\documentclass[5p]{elsarticle}

\usepackage{amsmath}
\usepackage{graphicx}
\usepackage{dcolumn}
\usepackage{bm}

\begin{document}

\title{The blogosphere as an excitable social medium: Richter's and Omori's Law in media coverage}

\author[IIASA,MUW]{Peter Klimek}
\author[IIASA]{Werner Bayer}
\author[MUW,SF,IIASA]{Stefan Thurner\corref{cor1}}
\cortext[cor1]{Corresponding author, stefan.thurner@meduniwien.ac.at}

\address[IIASA]{IIASA, Schlossplatz 1, A 2361 Laxenburg; Austria;}
\address[MUW]{Section for Science of Complex Systems, Medical University of Vienna,  
Spitalgasse 23, A 1090 Vienna; Austria;}
\address[SF]{Santa Fe Institute; 1399 Hyde Park Road; Santa Fe; NM 87501; USA} 


\begin{abstract}
We study the dynamics of public media attention by monitoring the content of online blogs.
Social and media events can be traced by the propagation of word frequencies of related keywords.
Media events are classified as exogenous -- where blogging activity is triggered by an external news item -- 
or endogenous where word frequencies build up within a blogging community without external influences. 
We show that word occurrences  show statistical similarities to earthquakes. The size distribution of media events 
follows a Gutenberg-Richter law, the dynamics of media attention before and after the media event follows Omori's law.
We present further empirical evidence that for media events of endogenous origin the overall public reception 
of the event is correlated with the behavior of word frequencies at the beginning of the event, and is to a certain degree predictable.
These results may imply that the process of opinion formation in a human society might be related to effects known from excitable media. 
\end{abstract}

\maketitle

\section{Introduction}

Modern man is exposed to a constant stream of news which we resorb, read, digest, discuss, disseminate and forget.
Most news items are of relatively little impact.
They receive a small amount of public attention over a short time and quickly descend into public oblivion.
However, occasionally news reports have a massive impact so that they can overthrow public opinions, heralded beliefs and even governments.
A recent example are the 2010-2011 Tunisian protests being sparked by reports of police use of tear gas against young demonstrators.
Of course, the reasons for these protests go far beyond this single incident, they involve multiple political, social and economical dimensions.
It is remarkable how reports about such relatively small incidents propagate under certain circumstances through a society -- 
i.e. a strongly interconnected system -- and trigger a nation-wide revolution, while under under circumstances the news will practically 
vanish unheard. It is tempting to see such media events as a human, social excitable medium.
One may view them as a social analog to earthquakes \cite{Gutenberg44, Kagan91}.
External stimuli trigger relaxation events where accumulated ``energy'' is discharged or spread 
within a complex, networked system.
This phenomenon can also be observed in other excitable media such as the brain \cite{Osorio10}, oscillating chemical reactions like the Belousov-Zhabotinsky reaction \cite{Belousov59} or the Mexican wave (or {\it La Ola}) at sport events \cite{Farkas02}.

Are there quantitative patterns in the way societies react to the arrival of breaking news?
How can  the impact of a news items be quantified?
How long and with which intensity do people devote their attention to current news?
Such research questions become amenable to quantitative study through online news discussions.
In recent years weblogging (or blogging) has emerged as a new publishing medium at the grassroots level of society. 
A blog is usually defined as a web page with entries listed in reverse chronological order, maintained by one or several writers. 
Blogs are often devoted to a unique topic, such as politics, finance or sports, and provide news or commentaries,
often several times per day.
In fall 2007 the blog search engine technorati.com stopped tracking the number of active blogs once it exceeded 100 million.
This rapid development sparked academic interest. The dynamical evolution of the {\em blogosphere}, defined as the collection of all blogs at a given time, 
has been  studied from a network perspective \cite{Kumar05}, where nodes are blogs which are connected if one blog 
possesses a hyper-link or URL to the other one. 
The spreading of news items can be seen as a diffusion process on this network \cite{Gruhl04}.
The accelerating shift of human conversations and discussions toward online media like blogs, twitter, etc.  
has reached a point where web data mining techniques \cite{Liu06} can be used to e.g. predict 
spikes of sales ranks of books \cite{Gruhl05}, measure public sentiments towards societal issues \cite{Liu07} or predict stock market movements \cite{Zhang09}.

In this work we study the impact of news reports on a small segment of the blogosphere
 by looking at timeseries of word frequencies. 
The sample contains blogs covering US domestic politics over a period of nearly two years.
Blogs are typically not the first to report a story.
They pick up developing news topics and offer comments and subjective points of view.
They are therefore an ideal candidate to measure how the public 'digests' news.
We show that the public reception of news reports follow a similar statistic as earthquakes do.
The intensity of fore- and aftershocks can be described by a power law in analogy to Omori's law \cite{Utsu95},
the size distribution of media events follows a Gutenberg-Richter law \cite{Knopoff82}.
It has been reported previously that Omori's law holds for the round-trip times of data packages in the internet \cite{Abe03}.
Power law signatures in human activity have recently been found in the distributions of waiting times between
a catastrophe and humanitarian responses \cite{Crane10} or between contributions to a discussion in online forums \cite{Mitrovic10}.


\section{\label{sec:data}Data and Methods} 

A dataset of 168 large and massively  popular blogs devoted to US domestic politics  covering the entire political spectrum was compiled.
The blog authors include popular political journalists (e.g. Glenn Beck or Taylor Marsh) as well as a number of
self-pronounced political commentators labeling themselves everything from 'far right' to 'liberal curmudgeons'.
We recorded the content of each blog entry with a time stamp with a resolution of one second. 
We focus on the time period of 670 days between July 1$^{\mathrm{st}}$ 2008 and May 3$^{\mathrm{rd}}$ 2010.
As the dominant themes this period  contains 
the 2008 US presidential elections, the advent of Sarah Palin, the health care reform and the Iraq torture scandal .
We developed a proprietary web crawler that continually crawls specified websites. 
The crawler can be targeted to a web page of a blog containing a blog entry of a given date.
The content of this site is automatically parsed and checked for hyper-links to other blog entries.
If a new link is found the linked page is downloaded and stored.
The software then analyzes the structure of the blog entry, identifying and storing its date, headline, and entire 
content. 
Once all links in the web page containing the blog entry have been identified the initial blog page is searched for a link to another web page containing older
entries of the same blog. This procedure is repeated as long as older entries are found.
The crawler is implemented in the Java programming language in a fully object oriented fashion. 
All collected data is stored in a SQL database allowing efficient subsequent sorting and filtering.
Filtering and pre-precessing was done  by removal of all short high frequency words e.g. 'the' or 'and',  and removal of 
plural-s and other frequent endings such as 'ing' or 'ed'. 

We will refer to words surrounding political topics actively discussed in the blogosphere by 'keywords'.
They were extracted by the following procedure.
The frequency of all $26^3=17576$ possible letter triplets (that is 'aaa', 'aab', $\dots$, 'zzz') was counted over all blog entries
for each day.
Nearly 50\% of these triplets where discarded since they never occurred.
From the remaining triplets we extracted the time series of daily frequencies and looked at their co-occurrences with words.
In particular, for each remaining triplet we searched for the day with the highest frequency, assume this day is $t_0$.
For this day a list of all words containing the triplet was made.
For each word from this list the time series of daily word frequencies in the range between thirty days before and after $t_0$ was extracted.
The word was kept for further analyses only if its daily word frequency has a maximum value within this range at $t_0$.
This procedure left us with approximately 4000 keywords.

\begin{figure}
\begin{center}
\includegraphics[height=4.0cm]{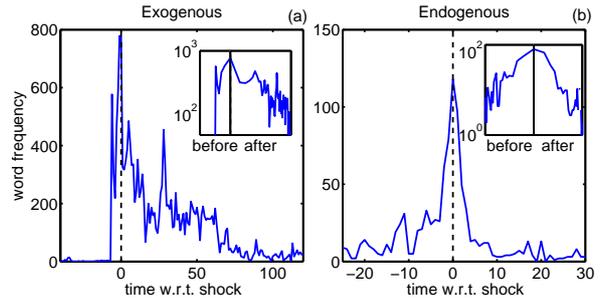}
\end{center}
\caption{{\it (a):} Word frequencies of 'palin' before and after the nomination of Sarah Palin as vice presidential candidate in 
the 2008 presidential elections. The $x$-axis is time (days) before and after the event, the event itself is indicated by a vertical 
dashed line. Note the absence of pre-cursory activity. {\it (b):} Word frequencies of 'inauguration' 
before and after the inauguration of Barack Obama. There is pronounced pre-cursory growth. Insets: same in log-log scale. 
}
\label{exoendo}
\end{figure}

Let $w_i(t)$ be the {\em word frequency} of keyword $i$ at time $t$. 
An event is (somewhat arbitrarily) identified as a strong increase in word frequencies over a short period of time, with a subsequent decay to 
(usually) the previous levels. The time at which the peak occurs within an event is $t_0$.
 The {\it event-size} $E$ is defined  by the peak level of word frequency relative to the average level within a timespan $T$  before and after the peak level 
 during the event.
We work with $T=30$. The event-size for word $i$ at peak-time $t_0$ is
\begin{equation}
E_{i,t_0} = \frac{1}{2 T +1} \frac{w_i(t_0)}{\sum_{t=t_0-T}^{t_0+T} w_i(t)} \quad.
\label{eventsize}
\end{equation}
 Two classes of response functions in social systems have been previously identified: 
endogenous and exogenous, see e.g.  \cite{Crane08}.
Endogenous events start with a phase of pre-cursory growth, exogenous ones by a sudden burst in word frequencies.
Both types of events are followed by a relaxation process after the peak is reached. It usually follows a  power law.
 
An example for an exogenous event is the nomination of Sarah Palin as vice presidential candidate
in the 2008 US presidential elections. Fig. \ref{exoendo}(a), shows the word frequencies of 'palin' before and after her nomination.
The dashed line indicates the peak in word frequency, a sudden jump is followed by a relaxation process.
As an example for an endogenous process consider the word frequencies of 'inauguration' before and after the inauguration of Barack Obama, Fig. \ref{exoendo}(b).
The day of the inauguration itself here coincides with $t_0=0$, the $x$-axis shows the days before and after the inauguration.
There is a phase of pre-cursory growth culminating at $t_0$  which is followed by a relaxation process.

\begin{figure}
\begin{center}
\includegraphics[height=5cm]{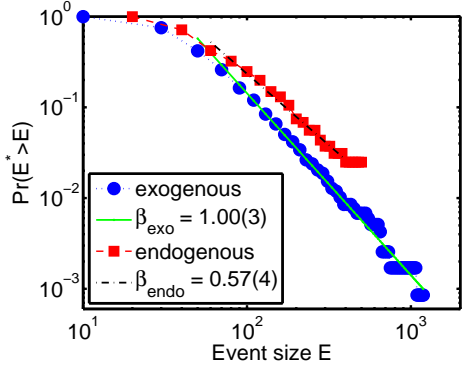}
\end{center}
\caption{Cummulative distribution  $\mathrm{Pr}(E^*>E)$ of event size $E$ for exogenous and endogenous events. 
The exogenous case (blue circles, dotted line) can be fitted by a Gutenberg-Richter law with exponent $\beta_{exo} = 1.00(3)$ (green solid line).
The exponent for the endogenous event sizes (red squares, dashed line) is $\beta_{exo}=0.57(4)$ (black dash-dotted line), 
close to the typical value for earthquakes.}
\label{CDF} 
\end{figure}

It has been suggested \cite{Crane08} that word frequencies before and after an endogenous event at time $t_0$ follow power laws of the form  
\begin{eqnarray}
w_i(t<t_0) & \propto & \left(t_0-t\right)^{-\alpha_{g}} \quad \nonumber \\
w_i(t>t_0) & \propto & \left(t-t_0\right)^{-\alpha_{d}} \quad,
\label{EndoEq}
\end{eqnarray}
where in general the  growth exponents before the peak at $t_0$,  $\alpha_{g}$, and the decay exponent after $t_0$, $\alpha_{d}$, can be different.
For the exogenous case there is no clear functional form for the word frequencies before $t_0$, the relaxation
dynamics often is of power law type \cite{Crane08},
\begin{equation}
w_i(t>t_0) \propto \left(t-t_0\right)^{-\gamma} \quad.
\label{ExoEq}
\end{equation}
Exogenous events -- translated into a seismological language -- can be identified with the Omori law where $w_i(t>t_0)$ is interpreted as 
the analog to the aftershock rate. 
The idea is that the frequencies of mentionings of a keyword
before (after) $t_0$ can be regarded as foreshock (aftershock) rates of that particular event.
For endogenous events we also observe 'foreshocks' following an Omori law.

For each event in a timeseries of word frequencies in our database it was checked whether it is endogenous or  exogenous. 
For this we segmented each timeseries (often containing several events) into windows of $T$ days and located the local maxima within this timespan
\footnote{To fit a power growth or decay we required at least non-zero word frequencies over two weeks before and after $t_0$.
The range of the fit was chosen between $(t_0, t_0 \pm \tau)$ with $\tau \in \{14, \dots, 30\}$.
For each value of $\tau$ in this range the Akaike Information Criterion was computed.
The exponent of the fit with minimal value of this criterion was used for further analyses.}.
The event was classified as endogenous or exogenous if the Residual Sum of Squares for the fit was below 0.15, 
computed by fits to Eqs. (\ref{EndoEq}) and (\ref{ExoEq})  around $t_0$, and  normalized by $w_i(t_0)$.

\section{\label{sec:results}Results}

With the above procedure we ended up with approximately 150 endogenous and 1000 exogenous events, i.e.  
an average of $0.2$ endogenous and $1.5$ exogenous events per day. Note that exogenous events are about an order of magnitude more frequent.
One media event generally corresponds to events in several word frequencies of related keywords.
This effect should be the same for endogenous and exogenous types.

The Gutenberg-Richter law is an empirical power law describing the frequency of earthquakes of a given radiated seismic energy $E_s$, 
\begin{equation}
\mathrm{Pr}(E_s^*>E_s) \propto E_s^{-(\beta+1)} \quad,
\label{GRL}
\end{equation}
with a typical exponent of $\beta \approx 2/3$ \cite{Utsu95}. 
Figure \ref{CDF} shows the cumulative distribution function (CDF) $\mathrm{Pr}(E^*>E)$ for endogenous and exogenous events. In both cases we find a Gutenberg-Richter law
\begin{equation}
\mathrm{Pr}(E^*>E) \propto E^{-(\beta+1)} \quad.
\label{GRLendoexo}
\end{equation}
For endogenous events we find  $\beta_{endo} = 0.57(4)$, for exogenous $\beta_{exo} = 1.00(3)$.
Together with the relative high number of exogenous compared to endogenous events, this suggests that exogenous events
share characteristics of 'earthquake swarms' \cite{Sykes70} -- sequences of small earthquakes over a relatively short time period.

\begin{figure}
\begin{center}
\includegraphics[height=4.5cm]{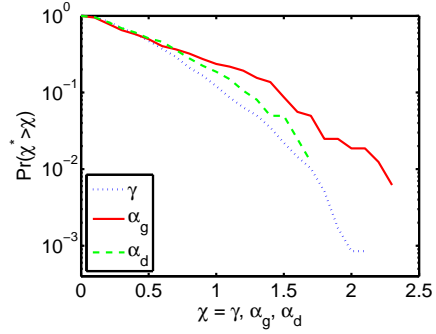}
\end{center}
\caption{Cumulative distribution functions of $\gamma$ (blue dotted line), $\alpha_{g}$ (red solid) 
and $\alpha_{a}$ (green dashed). The similar curves for $\gamma$  and  $\alpha_{d}$ suggest a 
universal decay law of public attention for media events. 
The exponents for pre-cursory growth $\alpha_{g}$ are typically higher than the decay exponents $\alpha_{d}$. }
\label{HistExp}
\end{figure}

The cumulative distribution functions for $\gamma$ for exogenous events and  $\alpha_{d}$ and $\alpha_{g}$ for endogenous events 
are shown in Fig. \ref{HistExp}. 
The value of the CDF of $\gamma^*$ at any given value $\gamma$ is defined as the probability that $\gamma^*$ is greater than $\gamma$, 
that is $\mathrm{Pr}(\gamma^* \geq \gamma)$.
Exponents $\gamma$ and $\alpha_{d}$ follow almost the same distribution function. 
This might suggest a universal decay law of public attention to media events.
For high values of exponents the CDF of growth exponents $\alpha_{g}$ is larger than the distribution function for decay
exponents.
In general we find for a given exponent value $\chi$
\begin{equation}
\mathrm{Pr}(\chi > \alpha_{g}) > \mathrm{Pr}(\chi > \alpha_{d}) \approx \mathrm{Pr}(\chi > \gamma)
\quad.
\label{exponents}
\end{equation}

\begin{figure}
\begin{center}
\includegraphics[height=4.5cm]{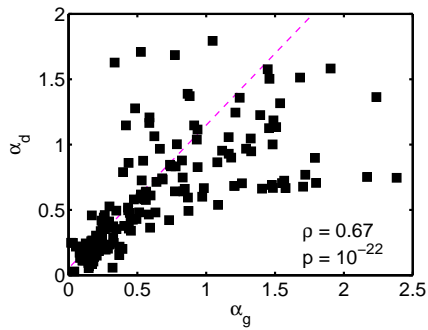}
\end{center}
\caption{Scatter plot of $\alpha_{g}$ versus $\alpha_{d}$. Each point is one endogenous event with $x$- and $y$-coordinates given by the values of the $\alpha$'s in Eqns. \ref{EndoEq}. The dashed line is a regression line, we find a correlation coefficient of $\rho \approx 0.67$ and a $p$-value of $10^{-22}$ against the null hypothesis of no correlation.}
\label{alpha} 
\end{figure}

To what extent can one predict the dynamics of endogenous events from their pre-cursory growth?
Fig.\ref{alpha} shows for each endogenous event its growth $\alpha_{g}$ versus its decay exponent $\alpha_{d}$.
Especially for low values of the exponents (smaller than e.g. one) there is a clear correlation between them.
Even if we include the entire dataset of endogenous events we can reject the null hypothesis of no correlation up to
a $p$-value of $10^{-22}$. The correlation coefficient is $\rho \approx 0.67$. 
This correlation should drastically increase if one would only include 
events with small growth and/or decay exponents $\alpha$.

\section{\label{sec:concl}Conclusions} 

Blogs offer a new and exciting possibility to study collective human behavior on a quantitative basis.
The blogosphere is a highly connected virtual space where people with different background and attitude
disseminate and discuss issues that caught their attention and sufficient interest, in a way that can be exploited for
quantitative studies.
We empirically studied how people react to new pieces of information through dynamical patterns of their blogging behavior.
We see the blogosphere is a human, excitable social medium in which waves of collective excitement can be traced. 
We studied event-size, foreshock and aftershock distributions in this medium and noticed analogies to seismology.
The intensity of fore- and aftershocks follows Omori's law, the distribution of event-sizes is of Gutenberg-Richter type.
We presented indications that there exist significant correlations between the dynamics of fore- and aftershocks.
One might also think of a 'Richter scale' for media events.
The largest event recorded in our dataset is the nomination of Sarah Palin as vice presidential candidate. 
Indeed, aftershocks of this event are still trembling and quivering through our society.

\end{document}